\documentclass[preprint, 12pt]{elsarticle}

\usepackage{amssymb}

\usepackage{amsmath}

\usepackage{multirow, array} 
\usepackage{float} 
\usepackage{graphicx}

\usepackage{amsfonts}
\usepackage{tikz}

\usepackage{cite}

\usepackage{cite}
\usepackage{tikz}
\usepackage{cuted}
\usepackage{rotating}
\usepackage{comment}
\usepackage{xcolor}

\newcommand{\be}{\begin{equation}}
\newcommand{\ee}{\end{equation}}
\newcommand{\bea}{\begin{eqnarray}}
\newcommand{\eea}{\end{eqnarray}}

\usepackage{color}

\journal{XXX}

\begin{document}

\begin{frontmatter}



\title{Kinetic Parameters analysis of GdAlO$_3$ based on thermoluminescent phenomenon}

\author[CICATA]{D. Nolasco-Altamirano}
\author[UAS]{C.S. Romero-Nuñez}
\author[UMich]{A. Alonso-Sotolongoza}
\author[CIEMAT]{J.F. Benavente}
\author[CICATA]{R. García-Salcedo}
\author[INNN]{O.A. García-Garduño}
\author[UMich]{J. Zarate-Medina}
\author[CIEMAT]{V. Correcher}
\author[CICATA]{T. Rivera Montalvo}

\address[CICATA]{Centro de Investigación en Ciencia Aplicada y Tecnología Avanzada del Instituto Politécnico Nacional, Legaria 694, 11500, Ciudad de México, México.}
\address[UAS]{Universidad Autónoma de Sinaloa, 80030 Culiacán Sin. México.}
\address[UMich]{Universidad Michoacana de San Nicolás de Hidalgo, 58000 Morelia Mich., México.}
\address[CIEMAT]{CIEMAT, Av. Complutense 40, 28040, Madrid, Spain}
\address[INNN]{Unidad de Radioneurocirugía, Instituto Nacional de Neurología y Neurocirugía, Insurgentes Sur 3877, Ciudad de México, México.}

\date{\today}

\begin{abstract}
We herein report on the calculation of thermoluminescence (TL) kinetic parameters determined from the TL emission of synthetic GdAlO3 (GAO) phosphors prepared by the co-precipitation method. The sample, characterized by means of X-ray diffraction with an orthorhombic phase structure (space group Pnma (62), shows complex glow curves consisting of at least four groups of components peaked at 100, 140, 240, and 290 °C where the two lower overlapped temperature peaks are difficult to identify using the TM-Tstop. The coexistence of both a continuum in the trap distribution (linked to the lower temperature peaks) and a discrete trap system (associated with the components at temperatures higher than 200 °C) can be distinguished. The estimation of the TL kinetic parameters is performed using GlowFit, TLAnal, the spreadsheet Origin, Computing Glow Curve Deconvolution (CGCD), and various heating rate (VHR) methods. However, only CGCD appears as the suitable technique for such purpose since it provides information on the TL physical process supported by mathematical models based on a linear combination of functions related to the First Order Kinetic approach.
\end{abstract}

\begin{keyword}
GdAlO$_3$, XRD \sep Thermoluminescence \sep kinetic parameters \sep HR \sep $T_m-T_{stop}$ 
\end{keyword}

\end{frontmatter}

\section{Introduction}

Gadolinium aluminate (GdAlO$_3$, GAO-) is a ceramic material that falls within the ABO$_3$ perovskite system, renowned for its remarkable optical, mechanical, and stable physical-chemical properties \citep{zhang2021structural, idrissi2021structural, jisha2015facile}. Given these attributes, GAO finds versatile applications, serving as an optical matrix material \citep{sajwan2017recent, kumar2016spectral, brylew2014studies, matos2011synthesis, tiwari2019synthesis}, scintillator \citep{qi2020hydrothermal, jisha2017luminescent}, thermometric material, and reinforcement material in composite materials \citep{girish2017supercritical, delice2019low, sotolongo2024influence, chen2018synthesis}. Furthermore, its luminescent properties make it a promising candidate for potential use in dosimetry \citep{sotolongo2022thermoluminescence, nolasco2022thermoluminescent}. GAO exhibits various types of defects, including intrinsic, extrinsic, and structural, all of which play pivotal roles in luminescence.

Thus, intrinsic defects within the crystal lattice of GAO, such as vacancies, interstitials, and antisites, play a crucial role in thermoluminescence by acting as traps for charges induced by ionizing radiation. This intimate relationship between the material's structure and its luminescent behavior is evident in how these defects capture the charges generated by radiation, holding them until they recombine with holes within the material. Understanding this trapping and recombination dynamics is essential for comprehending the light emission in the thermoluminescent process, thus directly linking GAO's crystal structure and its luminescent response.

Intrinsic defects within the crystal lattice of GAO result from deviations from ideal stoichiometry, manifesting as vacancies (absence of Gd and Al sublattices), interstitials (presence of atoms occupying interstitial sites within the GAO lattice, potentially involving Gd, Al, or O atoms), and antisites (e.g., the substitution of Gd atoms for Al atoms or vice versa, leading to local disorder within the crystal structure) \citep{tamrakar2016studies, chaudhury2007high}. Extrinsic defects, introduced during GAO synthesis, typically include impurities, dopants, and vacancies resulting from non-stoichiometric composition. Additionally, structural defects deviate from the ideal GAO structure, primarily comprising grain boundaries (interfaces between neighboring crystalline grains in polycrystalline GAO samples, potentially acting as barriers to electron transport), dislocations (deviations from the ideal crystal lattice structure occurring during crystal growth or as a result of mechanical deformation), and stacking faults (deviations from the regular stacking sequence of atomic planes within the crystal lattice) \citep{quirk2024grain, isotta2023microscale, hu2017grain}. These defects serve as traps for the negative charges induced by ionizing radiation, which remain trapped until recombining with holes, also possibly hosted within defects within the material \citep{nowotny2015defect, zafar2022recent}.

The electron-hole recombination results in radiative emission, leading to TL emission. The wavelength of the photon emission depends on the type of defect hosting each trapped charge, as different defects may possess varying activation energies ($E_a$) for emission. These kinetic parameters, including the activation energy ($E_a$), frequency factor ($s$), and kinetic order ($b$), can be mathematically determined using various methods based on the shape of experimental TL curves, peak position, and glow intensity \citep{correcher2024thermoluminescence, bos2017thermoluminescence}. However, each method employed for estimating TL kinetic parameters has its own set of advantages and limitations. 

GlowFit provides a user-friendly interface for fitting TL curves and offers flexibility in selecting fitting models and parameters \citep{bakr2020determination}. However, it may lack advanced features for handling complex TL glow curves with overlapping peaks and is limited in handling non-standard TL glow curve shapes. TLAnal allows comprehensive analysis capabilities for TL glow curves, including deconvolution and parameter determination; however, it is not user-friendly due to the complexity of the software interface and analytical techniques \citep{benavente2019thermoluminescence, benavente2020numerical, benavente2020characterization}. The spreadsheet Origin enables customized data analysis and visualization using a familiar spreadsheet interface and extensive plotting and graphing capabilities for presenting TL data. However, it is limited in performing complex mathematical modeling and parameter fitting compared to dedicated TL analysis software, and it does not consider the physical processes leading to the TL mechanism \citep{sadek2015deconvolution, sadek2015properties}.

The Various Heating Rate (VHR) method allows the investigation of kinetic parameters under different heating rate conditions, offering insights into the underlying TL mechanisms \citep{majgier2021application}. However, it requires careful experimental design and data interpretation to account for heating rate effects. It may give rise to wrong estimations of TL kinetic parameters due to experimental artifacts or uncertainties in heating rate calibration. Peak Shape provides information on the distribution and characteristics of TL peaks, such as peak width, symmetry, and asymmetry. It allows the identification of peak components and peak purity assessment \citep{kitis2007peak}. However, the interpretation of peak shape may be subjective and influenced by experimental factors, and it may require complex mathematical modeling and analysis to extract information from complex peak shapes. Finally, Computing Glow Curve Deconvolution (CGCD) is designed for deconvolution and analysis of TL glow curves, particularly those with overlapping peaks. It provides well-defined algorithms for extracting kinetic parameters from complex TL data \citep{benavente2020numerical, wazir2022computerized, mckeever1985thermoluminescence}. Although this method requires expertise in TL analysis techniques and software operation, CGCD is the most suitable method for assessing TL kinetic parameters since it supports the physical processes producing TL glow peaks with mathematical models based on a linear combination of functions.

This study analyzes the TL kinetic parameters derived from the TL emission of synthetic GdAlO$_3$ phosphors synthesized via the co-precipitation method. These TL kinetic parameters are estimated using GlowFit, TLAnal, the spreadsheet Origin, CGCD, and VHR methods.

\section{Materials and Methods}

\subsection{Synthesis and Structural Characterization}

The co-precipitation method was employed to synthesize GAO powders. Starting materials included Al(NO$_3$)$_3 \cdot$9H$_2$O (98.7\% J.T. Baker) and Gd$_2$O$_3$ (99.9\%, Alfa Aesar), which were weighed in stoichiometric ratios to ensure proper stoichiometry. The precipitation occurred under constant stirring and pH control using ammonia water. The precursor underwent washing and filtration following a specific duration until gel formation was achieved. Subsequently, the gel underwent calcination at various temperatures.

The synthesized powders underwent X-ray diffraction (XRD) analysis in the second stage. X-ray diffraction patterns were obtained using a D8 Advance device from Bruker, utilizing Cu–K$alpha$ radiation with a wavelength of 1.541. The analysis spanned a 2$theta$ range from 20° to 70° with a time step of 0.6 s.

TL glow curve emission was recorded using a Lexsyg Smart TL/OSL reader system from Freiberg Instruments, featuring an internal $^{90}$Sr/$^{90}$Y beta source with a dose rate of 110 mGy/s. Prior to TL measurements, powders were weighed at 40 $\mu$g per aliquot, with two aliquots used for each measurement. Experimental data collection involved linear heating at a rate of 2°C/s from room temperature up to 350°C in a nitrogen atmosphere, except for variations in heating rates during the VHR experiment.

\subsection{Thermoluminescence glow curve analysis}

The TL glow curves were investigated through a series of TL methods summarized as follows.

\subsubsection{$T_m-T_{stop}$ Method}

Two methods exist to separate overlapping TL peaks that occur due to the proximity of a series of single trap levels or continuously distributed trap levels. The first method involves a pre-heating treatment up to a stopping temperature ($T_{stop}$) before detecting a TL curve \citep{puchalska2006glowfit, majgier2021application}. The second method is based on exciting samples at different temperatures ($T_{exc}$) and recording the TL curve between starting and ending temperatures \citep{sotolongo2024influence}. This approach allows for the identification of additional peaks beneath a complex peak in the TL glow curve. These changes may include shifts in the position of the emission peaks, changes in the intensity of the peaks, and alterations in the overall shape of the curve. Studying these changes 
can identify and characterize the different TL emission peaks, allowing them to better understand the underlying properties and processes in the material under study.

Experimental data were collected using 13.2 Gy beta irradiation with a heating rate of 10 °C/s until the $T_{stop}$ temperature was reached. The method was repeated to generate a plot of $T_{M}$ \textit{vs}. $T_{stop}$, increasing $T_{stop}$ from 50°C to 270°C in 10°C increments. Moreover, employing these methods provided specific information about the TL material characteristics, such as the number of contributions and the characteristics of each distribution (localized or continuous trapped electron distributions).

\subsubsection{Glow Curve Computational Methods}

In recent decades, researchers have widely employed various Glow Curve Computational Methods (GCCM) tools to fit experimental measurements and to isolate individual glow peaks in the TL glow curve using the deconvolution method \citep{brylew2014studies, kitis2007peak, wazir2022computerized}. These tools are instrumental in obtaining the kinetic parameters for each peak comprising the entire TL curve. 

GCCM tools use various algorithms to decompose TL glow curves into their individual components. Among the most commonly used algorithms are curve fitting methods, which model the shape of the brightness curve using mathematical functions such as the least squares method, the weighted least squares method, and the nonlinear regression method.

The methodologies utilized can be summarized as follows. Below are some of the GCCM utilized in this study.

\paragraph{GlowFit}

In this study, the GlowFit analysis software, introduced in 2006 \citep{wazir2022computerized}, was utilized for fitting TL glow curves of GAO. The curve fitting methods employed by GlowFit include techniques such as the least squares method, weighted least squares method, and nonlinear regression. These methods aim to find the optimal combination of kinetic model parameters that minimize the difference between the experimental TL curves and those predicted by the model.

This software was developed for glow curve analysis based on the first-order kinetics model proposed by Randall and Wilkins \citep{randall1945phosphorescence}, which describes TL emission with the equation (\ref{RWEq}):
\begin{strip}
\vspace{2cm}
\be
    I(T) = I_M \left( \frac{E}{kT_M} - \frac{E}{kT} \right) \exp \left\{ \frac{E}{kT_M} \left[ \alpha \left( \frac{E}{kT_M} \right) - \frac{T}{T_M} \exp \left( \frac{E}{kT_M} - \frac{E}{kT} \right) \alpha \left( \frac{E}{kT} \right) \right] \right\}, \label{RWEq}
\ee
\end{strip}
where $I(T)$ represents the intensity of TL emission at temperature $T$, $I_M$ is the maximum intensity of TL emission, $E$ is the trap centers' activation energy, $k$ is the Boltzmann constant, $T_M$ is the temperature associated with the maximum emission intensity, and $\alpha(x)$ is a function with constant values generated by the program, defined by:
\be
    \alpha(x) = 1 - \frac{a_0 + a_1x + a_2x^2 + a_3x^3 + x^4}{b_0 + b_1x + b_2x^2 + b_3x^3 + x^4}. \label{Eq2}
\ee

This equation describes the TL emission rate as a function of temperature and other kinetic parameters related to trap centers in the material. Therefore, this method compares the kinetic parameters obtained with those derived from other methods.

\paragraph{TLAnal}

TLAnal, a program based on the general order model \citep{bakr2020determination, puchalska2006glowfit, chung2005computer}, provides a robust fitting procedure. This software, referred to as the General Approximation model, operates under the assumption that the free carrier concentration in the conduction band and its rate of change are significantly smaller than those of the trapped carrier concentration. Two minimization methods were employed for fitting: the Hessian and simplex methods \citep{mckeever1985thermoluminescence}.

The Hessian method utilizes information about a function's second derivatives to rapidly converge to an optimum \citep{hestenes2012conjugate}. In contrast, the simplex method broadly explores the search space and does not require derivative information \citep{maros2002computational}. Both methods are widely employed in optimization and curve-fitting problems.

\paragraph{TLOrigin}
The TLOrigin algorithm is designed for use within Microsoft Excel spreadsheets. It employs various curve-fitting methods, including weighted least squares and nonlinear regression, alongside optimization techniques to determine optimal general order kinetics parameters, as outlined in the equations proposed for the OTOR (one trap, one recombination center) model \citep{kitis1998thermoluminescence}. Additionally, it incorporates deconvolution algorithms to separate overlapping peaks in the TL curve and identify different trap centers. This multifaceted approach ensures a robust and precise analysis of TL curves within the Excel environment. Kinetic parameters, such as maximum intensity, temperature of the TL glow peak, kinetic order, and activation energy, are determined based on user-defined criteria, such as the figure of merit (FOM) value. Achieving acceptable FOM values often requires multiple iterations. By leveraging the "Solver" add-on \citep{fylstra1998design}, the computational process is executed, resulting in the determination of kinetic parameters \citep{chung2005computer, kitis1998thermoluminescence}.

\paragraph{Computational Glow Curve Deconvolution}

The Computational Glow Curve Deconvolution (CGCD) application was developed by JF. Benavente \citep{benavente2019thermoluminescence}, utilizes mathematical models grounded in a linear combination of functions associated with the First Order Kinetic approach \citep{benavente2019thermoluminescence, kazakis2019tldecoxcel, fylstra1998design, lasdon1974nonlinear}. 
 The linear combination of related features method involves the weighted sum of several individual features, where each feature represents a specific contribution to the behavior observed in the data. These features may relate to different physical processes or phenomena contributing to the TL glow curve. By combining these functions in a linear manner, the method seeks to accurately model the shape and behavior of the TL glow curve, allowing the identification and characterization of the different components present in the sample. This software excels in fitting experimental measurements owing to its remarkable physical consistency \citep{abramowitz1968handbook, jisha2015facile}, as expressed by the equation (\ref{Eq3}):
\begin{strip}
\be
    I_{\text{TL}}^i (\text{a.u.}) = I_N \left( \frac{\int_{E_1}^{E_2} f(E) \cdot \exp\left(\frac{-E}{K_b T}\right) \cdot \exp\left\{\frac{-E_0}{K_b T} \cdot \left(\frac{T}{T_N}\right) \cdot \exp\left(\frac{E_0}{K_b T} - \frac{E}{K_b T}\right) \cdot \alpha\left(\frac{E}{K_b T}\right)\right\} \, dE}{\int_{E_1}^{E_2} f(E) \cdot \exp\left(\frac{-E}{K_b T}\right) \cdot \exp\left\{\frac{-E_0}{K_b T_N} \cdot \exp\left(\frac{E_0}{K_b T_N} - \frac{E}{K_b T_N}\right) \cdot \alpha\left(\frac{E}{K_b T_N}\right)\right\} \, dE}\right), \label{Eq3}
    \ee
\end{strip}
where $f(E)$ is the expression for the trapped electron density, which can be written in different forms depending on the distribution of traps. For a continuous trap distribution, $f(E)$ can be expressed as an exponential Gaussian shape:
\be
    f(E) = \frac{1}{\sqrt{2\pi}\sigma} \exp\left( -\frac{(E - E_0)^2}{2\sigma^2} \right), \label{Eq4}
\ee
or
\be
    f(E) = \frac{1}{\sigma} \exp\left( -\frac{E - E_0}{\sigma} \right), \label{Eq5}
\ee
where $E_0$ is the mean trap energy, and $\sigma$ is the standard deviation of the trap energies.

In the case of a localized trap distribution, $f(E)$ can be expressed using the Dirac delta function:

\be
    f(E) = \delta(E - E_0) \label{Eq6}
\ee
which indicates that the trapped electron density is concentrated at a single energy level $E_0$. This means that all trapped electrons have the same energy $E_0$, with no spread or variation in energy levels. In physical terms, this localized trap distribution represents a situation where only one type of trap has a specific, well-defined energy level $E_0$. This is in contrast to a continuous trap distribution where the trap energies can vary over a range.

All fittings are performed using the Levenberg-Marquardt algorithm \citep{levenberg1944method, marquardt1963algorithm, benavente2019thermoluminescence}, in order to obtain the kinetic parameters $E$, $s$, and $b$ from the original TL glow curve of the sample beta-irradiated at an absorbed dose of 13.2 Gy, using any of the previously mentioned mathematical models. The kinetic parameters obtained for each peak at this irradiation dose are presented in Table 1. To compare the quality of the computational fittings, we also measured the FOM according to equation \citep{balian1977figure}:
\be
    \text{FOM} [\%] = \sum_{i=1}^{n} \left| \frac{y_i - \hat{y_i}}{A} \right| \times 100 \%,  \label{Eq7}
\ee
where $y_i$ are the experimental values, $\hat{y_i}$ are the corresponding fitted values, $A$ is the area of the fitted glow curve, and $n$ is the number of fitted data points.

\subsubsection{Peak Shape Method}

Grossweiner \citep{grossweiner1953note} established the first peak shape approach for first-order peaks, defining:
\be
    E = 1.51 \frac{k T_m T_1}{\tau}, 
\ee
where $T_m$ is the temperature at maximum intensity, $T_1$ is the temperature at half maximum intensity on the low-temperature side, and $\tau = T_m - T_1$. Lushchik \citep{lushchik1956investigation} further developed this method, using the high-temperature half-width $\delta = T_2 - T_1$ for first-order peaks:
\be
    E = k \frac{T_m^2}{\delta},
\ee
and for second-order peaks:
\be
    E = 2k \frac{T_m^2}{\delta}. 
\ee

The method used here is based on the modification of the equation for Chen  \citep{chen1969glow}, which is used to verify trapping parameters within the crystal. The following shape parameters are determined using points depicted in Figure \ref{F1}: total half-intensity width ($\omega = T_2 - T_1$), high-temperature half-width ($\delta = T_2 - T_m$), and low-temperature half-width ($\tau = T_m - T_1$), where $T_m$ is the peak temperature at maximum intensity, and $T_1$ and $T_2$ are temperatures on either side of $T_m$ corresponding to half peak intensity. These parameters help calculate the kinetic parameters $E$, $s$, and $b$.

The order of kinetics ($b$) is determined by calculating the glow peak's symmetry factor ($\mu$), using the known shape parameters:
\be
    \mu_g = \frac{T_2 - T_M}{T_2 - T_1} \label{Eq8}
\ee

The activation energy ($E$) was calculated using the Chen equations, providing the trap depth in terms of $\tau$, $\delta$, and $\omega$. A general formula for $E$ was given by Chen \citep{chen1969glow} as follows:
\be
    E_\alpha = C_\alpha \left( \frac{k T_M^2}{\delta} \right) - b_\alpha (2 k T_M), \label{Eq9}
\ee
where
\be
\begin{aligned}
    C_\tau &= 1.510 + 3.0 (\mu_g - 0.42) & \quad b_\tau &= 1.580 + 4.2 (\mu_g - 0.42) \\
    C_\delta &= 0.976 + 7.3 (\mu_g - 0.42) & \quad b_\delta &= 0 \\
    C_\omega &= 2.520 + 10.2 (\mu_g - 0.42) & \quad b_\omega &= 1 \label{Eq10}
\end{aligned}
\ee

The symmetry factor of the geometric shape, $\mu_g = \frac{\delta}{\omega}$, indicates the order of the TL glow peak. If $\mu_g = 0.42$, the TL glow peak is first order; if $\mu_g = 0.52$, it is second order.

The frequency factor $s$ can be calculated using the following equation:
\be
    s = \frac{\beta E}{k T_M^2} \exp\left(\frac{E}{k T_M}\right) \left[1 + (b - 1) \Delta_M \right]^{-1} \label{Eq11}
\ee
where $\beta$ is the heating rate, $k$ is the Boltzmann constant, and $\Delta_M$ is the peak width parameter.

The Peak Shape Method is used to determine the kinetic parameters of thermoluminescent peaks by analyzing their shape. However, when the TL glow curves exhibit a high degree of overlap between peaks, the application of this method becomes problematic. The difficulty in distinguishing individual peaks compromises the precision in the identification of their characteristics and, therefore, the kinetic parameters. Consequently, the Peak Shape Method will be analyzed in this study solely for comparative purposes with other methods. While this method offers a straightforward approach to determining kinetic parameters based on the shape of TL glow peaks, it may not be suitable for curves with highly overlapping components. By including this method in the analysis, the study aims to highlight its limitations and contrast its effectiveness with more advanced techniques, ensuring a comprehensive evaluation of the kinetic parameters.

\begin{figure}[h]
    \centering
    \resizebox{9cm}{!} {
    \begin{tikzpicture}
    \draw[thick,->] (0,0) -- (9,0) node[right] {Temperature (K)};
    \draw[thick,->] (0,0) -- (0,6.5) node[above] {Intensity (a.u.)};

    \draw[thick] plot [smooth] coordinates {(1,0) (2,0.5) (3,2) (4,5) (5,2) (6,0.5) (7,0)};

    \draw[dashed] (4,0) -- (4,5) node[above] {$T_m$};
    \draw[dashed] (3,0) -- (3,2) node[above] {$T_1$};
    \draw[dashed] (5,0) -- (5,2) node[above] {$T_2$};

    \draw[dashed] (0,5) -- (4,5) node[left] {};
    \draw[dashed] (0,2) -- (3,2) node[left] {};
    \draw[dashed] (4,2) -- (5,2);

    \draw[<->] (3,2.2) -- (5,2.2) node[midway,above] {$\omega$};
    \draw[<->] (4,0.2) -- (5,0.2) node[midway,below] {$\delta$};
    \draw[<->] (3,0.2) -- (4,0.2) node[midway,below] {$\tau$};

    \node at (-1,5) [above right] {$I_m$};
    \node at (-1,2) [above right] {$I_m/2$};
    \label{F1}
\end{tikzpicture}
}
    \caption{Schematic representation of a thermoluminescence (TL) glow curve. The peak intensity ($I_m$) and half maximum intensity ($I_m/2$) are indicated, along with the temperatures at these points: $T_m$ (the temperature at maximum intensity), $T_1$ (the temperature at half maximum on the low-temperature side), and $T_2$ (the temperature at half maximum on the high-temperature side). The parameters $\tau$, $\delta$, and $\omega$ represent the low-temperature half-width, high-temperature half-width, and total half-width, respectively.}
    \label{F1}
\end{figure}
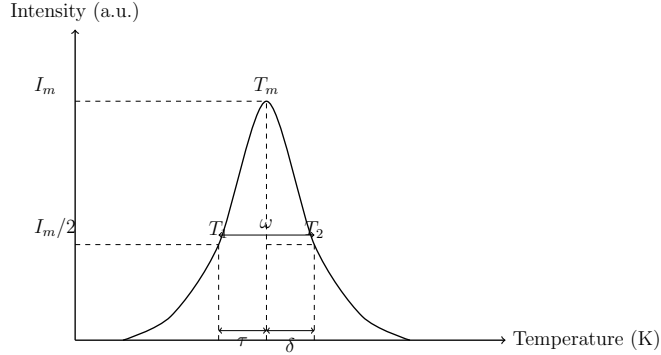

\subsubsection{Various Heating Rate Method}

The method of various heating rates VHR for evaluating the activation energy of a TL glow peak is based on the expression of first-order kinetics introduced by Randall and Wilkins \citep{randall1945phosphorescence}. This method relates the linear heating rate ($\beta$) to the maximum intensity in the TL glow peak. Although VHR methods were initially developed for first-order glow peaks only \citep{kitis2007peak}, they can be used to determine activation energy regardless of the kinetic order and yield good results in more general situations \citep{chen1969glow, chen2022various}. 

The position of the glow peak temperature $T_m$ is obtained by differentiating the Randall-Wilkins equation with respect to temperature $T$, given by:
\be
    \frac{\beta E}{k T_M^2} = s \exp \left( -\frac{E}{k T_M} \right). \label{Eq12}
\ee

Applying natural logarithms to the last equation, we can rewrite it as follows:
\be
    \ln \left( \frac{\beta}{k T_M^2} \right) = \ln \left( \frac{s}{E} \right) - \frac{E}{k T_M} \label{Eq13}
\ee

A linear relationship is established between $\ln \left( \frac{\beta}{k T_M^2} \right)$ and $\frac{1}{T_M}$, where the activation energy (in $eV$) of each peak can be obtained from the slope. This relationship has been demonstrated in TL glow curves measured using different heating rates \citep{majgier2021application}. 

The activation energy derived from $T_M$ for two different heating rates is given by \citep{mckeever1985thermoluminescence}:
\be
    E = k \left( \frac{T_{M1} \cdot T_{M2}}{T_{M1} - T_{M2}} \right) \ln \left[ \frac{\beta_1}{\beta_2} \left( \frac{T_{M2}}{T_{M1}} \right)^2 \right], \label{Eq14}
\ee
where $T_{M1}$ and $T_{M2}$ are the maximum peak temperatures corresponding to heating rates $\beta_1$ and $\beta_2$. The frequency factor $s$ can be calculated using the following equation:
\be
    s = \frac{E}{k} \exp \left\{ \frac{T_{M2} \ln \left( \frac{T_{M2}^2}{\beta_2} \right) - T_{M1} \ln \left( \frac{T_{M1}^2}{\beta_1} \right)}{T_{M1} - T_{M2}} \right\}. \label{Eq15}
\ee

\begin{figure}[h!]
    \centering
    \includegraphics[width=9cm]{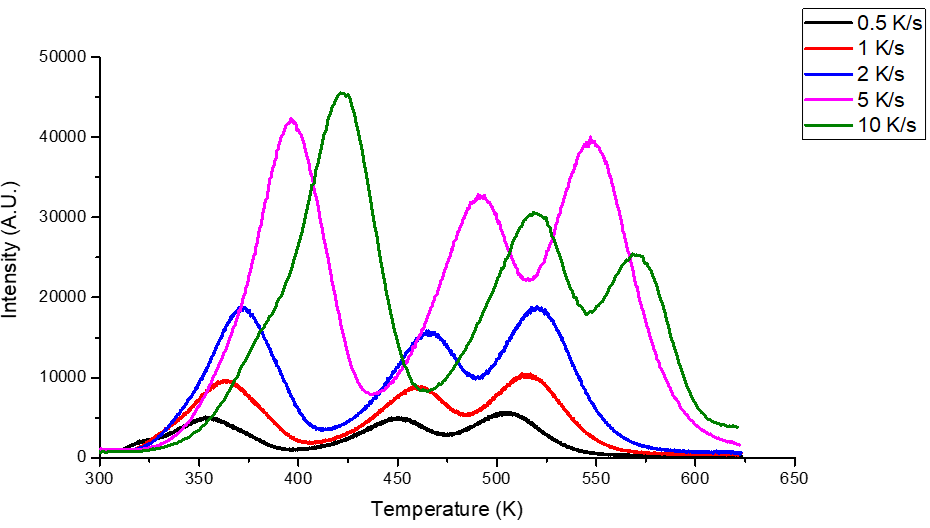}
    \caption{Thermoluminescence glow curves of GdAlO$_3$ after being exposed to 13.2 Gy beta radiation using different heating rate values (0.5 – 10 K/s).}
    \label{Fig1}
\end{figure}

To analyze TL glow curves using this technique, TL measurements were collected, and VHR experiments were conducted, as shown in Figure \ref{Fig1}. The TL glow curves were obtained with various VHRs in the range of 1°C/s to 20°C/s for 13.2 Gy beta irradiation. 

\section{Results}

\subsection{X-ray diffraction Analysis}

\begin{figure}[h!]
    \centering
    \includegraphics[width=8cm]{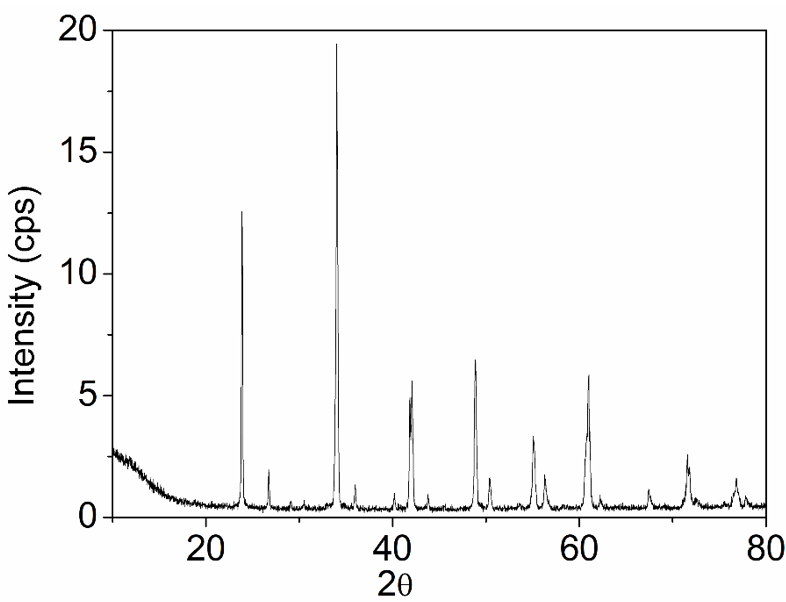}
    \caption{X-ray diffraction pattern of GdAlO$_3$ powders. The diffraction peaks correspond to an orthorhombic crystal structure in the Pnma (62) space group, consistent with JPCDS Card No. 48-0310. The narrow and sharp (132) and (231) peaks indicate good crystallization characteristics. }
    \label{Fig2}
\end{figure}

The structural characteristics of GdAlO$_3$ powders were analyzed using XRD, as shown in Figure \ref{Fig2}. The XRD pattern reveals that the sample exhibits a single-phase structure consistent with the standard JPCDS card no. 48-0310, corresponding to an orthorhombic crystal structure in the Pnma (62) space group. The prominent (132) and (231) diffraction peaks are narrow and sharp, indicating good crystallization characteristics of the prepared sample. These sharp peaks suggest a well-defined crystalline structure, confirming the successful synthesis of GdAlO$_3$ powders via the co-precipitation method.

\subsection{TL Glow Curve and Peaks Characterization}

\begin{figure}[h!]
    \centering
    \includegraphics[width=8cm]{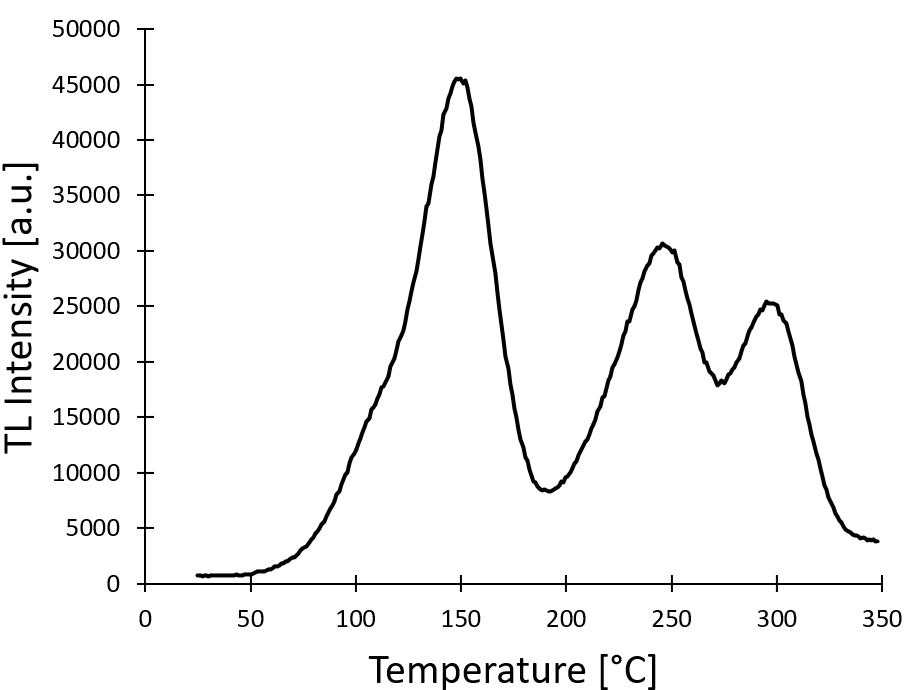} 
    \caption{TL glow curve of GdAlO$_3$ powders exposed to beta radiation at 13.2 Gy. The curve features a dominant peak near 140°C and two lower-intensity peaks at 246°C and ~300°C. The sample was heated up to 350°C at a rate of 10°C/s after beta irradiation.}
    \label{Fig3}
\end{figure}

The TL glow curve obtained from GdAlO$_3$ powders previously exposed to beta radiation at a dose of 13.2 Gy is presented in Figure \ref{Fig3}. The TL glow curve shows well-defined peaks under beta irradiation. The curve is dominated by a main TL peak near 140°C, with two additional low-intensity peaks observed at higher temperatures (246°C and approximately 300°C). The TL glow curve was collected by heating the sample up to 350°C at a rate of 10°C/s after beta irradiation.

A TL material can be characterized by analyzing its TL glow curves (Figure \ref{Fig3}). However, it is essential to use a coherent mathematical model that links the glow curves with the number and type of contributions (localized or continuous) and their corresponding kinetic approach. The $T_{Max}-T_{Stop}$ method becomes fundamental to achieve this.

\subsection{$T_{Max}-T_{Stop}$ results}

The $T_{Max}-T_{Stop}$ results, shown in Figure \ref{Fig4}, indicate the presence of at least three distinct contributions to the thermoluminescence glow curve. This means that the glow curve is composed of multiple overlapping peaks, each corresponding to different types of trapped electrons within the material.

The first contribution is associated with a continuous distribution of trapped electrons. This suggests that electrons are trapped in a broad range of energy levels, leading to a more spread-out glow peak.

The last two contributions, on the other hand, are compatible with localized distributions of trapped electrons. This implies that these peaks arise from specific, discrete energy levels where electrons are trapped, resulting in sharper, more defined peaks in the glow curve.

A mathematical model based on a linear combination of at least three peaks is employed to accurately fit the experimental TL data. This model follows the First-Order Kinetic approach, which helps identify and quantify the different contributions to the overall glow curve.

\subsection{Deconvolution Results}

The deconvolution method is one of the most efficient mathematical techniques for obtaining all kinetic parameters related to each contribution comprising the entire TL glow curve, especially when these contributions exhibit a high degree of overlap. A set of experimental TL glow curves was fitted using the deconvolution method, utilizing several software applications that implement both numerical methods and mathematical models. These applications include:

\begin{enumerate}
    \item \textbf{CGCD applications}. This uses a mathematical model based on a linear combination of different functions related to the First Order Kinetics approach, as follows in equation (\ref{Eq16}):
    \begin{strip}
    \be
        I_{TL}(T) = f^{Exp}\left(I_{Max}^{(I)}, T_{Max}^{(I)}, E_a^{(I)}, \sigma^{(I)} ;T\right) + f^{Gaus}\left(I_{Max}^{(II)}, T_{Max}^{(II)}, E_a^{(II)}, \sigma^{(II)} ;T\right) +  f^{Local}\left(I_{Max}^{(III)}, T_{Max}^{(III)}, E_a^{(III)} ;T\right) + f^{Local}\left(I_{Max}^{(IV)}, T_{Max}^{(IV)}, E_a^{(IV)} ;T\right) \label{Eq16}
    \ee
    \end{strip}
    where the mathematical functions are related to equations (\ref{Eq3}), (\ref{Eq4}), (\ref{Eq5}), and (\ref{Eq6}). Figure \ref{Fig5}(A) shows the results obtained using this deconvolution method.
    
    \item \textbf{GlowFit}. This free software implements a mathematical model based on a linear combination of five functions related to the General Order Kinetics approach. Figure \ref{Fig5}(B) shows the results obtained using this deconvolution method.

    \item \textbf{TLANal}. 
    This uses a mathematical model similar to the previous one, but the peak centered at 246°C is divided into two contributions. This division is justified by the shape of the TL glow curve, suggesting overlapping peaks that require separate components for a more accurate fit. Figure \ref{Fig5}(C) shows the results obtained using this deconvolution method.

    \item \textbf{SpreadSheet Origin software}. This uses a mathematical model similar to the previous ones. Figure \ref{Fig5}(D) shows the results obtained using this deconvolution method. The differences observed between the values of the kinetic parameters obtained when fitting experimental curves using deconvolution methods (Table \ref{tab:kinetic_parameters}) are within the range of the degeneration of the merit function used for the evaluation \citep{romero2023correlation}. These discrepancies will be amplified by using different numerical methods implemented by each application to perform said adjustment, whose common characteristic is their convergence to low FOM.
\end{enumerate}

\begin{figure}[h]
    \centering
    \includegraphics[width=8cm]{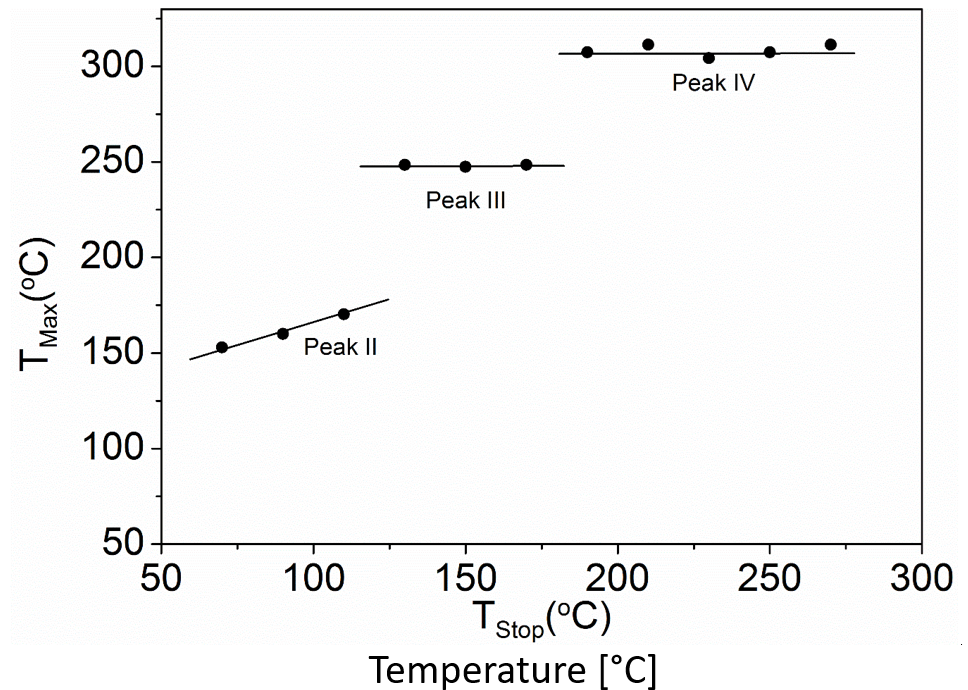} 
    \caption{$T_{Max}-T_{Stop}$ results of the glow curves of GdAlO$_3$ obtained using a heating rate of 2 K/s. The results show the presence of at least three distinct contributions: the first is related to a continuous distribution of trapped electrons, indicating a broad range of energy levels, while the last two are compatible with localized distributions, suggesting discrete energy levels.}
    \label{Fig4}
\end{figure}

\begin{sidewaystable}
\centering
\caption{Comparison of kinetic parameters: activation energy values ($E$) and frequency factor obtained from different methods for GdAlO$_3$.}
\resizebox{20cm}{!} {
\begin{tabular}{|c|c|c|c|c|c|c|c|c|c|c|c|c|}
\hline
\textbf{Methods} & \multicolumn{3}{|c|}{\textbf{Peak 1}} & \multicolumn{3}{|c|}{\textbf{Peak 2}} & \multicolumn{2}{|c|}{\textbf{Peak 3}} & \multicolumn{2}{|c|}{\textbf{Peak 4}} & \multicolumn{2}{|c|}{\textbf{Peak 5}} \\ 
\hline
 & $E$ (eV) & $s$ (s$^{-1}$) & $\sigma$ (eV) & $E$ (eV) & $s$ (s$^{-1}$) & $\sigma$ (eV) & $E$ (eV) & $s$ (s$^{-1}$) & $E$ (eV) & $s$ (s$^{-1}$) & $E$ (eV) & $s$ (s$^{-1}$) \\ 
\hline
\textbf{GlowFit}   & 1.160 & 9.11$\times$10$^7$ & - & 2.440 & 2.94$\times$10$^{17}$ & - & 2.820 & 6.36$\times$10$^{17}$ & 2.990 & 4.36$\times$10$^{17}$ & 1.150 & 1.49$\times$10$^6$ \\ 
\hline
\textbf{TLA}       & 0.663 & 1.40$\times$10$^8$ & - & 1.440 & 8.10$\times$10$^{14}$ & - & 1.117 & 5.12$\times$10$^{10}$ & 2.288 & 9.41$\times$10$^{19}$ & 2.226 & 3.38$\times$10$^{16}$ \\ 
\hline
\textbf{TLOrigin}  & 0.810 & 5.40$\times$10$^{10}$ & - & 1.260 & 4.12$\times$10$^{15}$ & - & 1.060 & 8.41$\times$10$^{10}$ & 2.000 & 1.33$\times$10$^{20}$ & 1.980 & 1.44$\times$10$^3$ \\ 
\hline
\textbf{Peakshape} $\tau$ & 0.836 & 5.15$\times$10$^{10}$ & - & 1.354 & 1.14$\times$10$^{16}$ & - & 1.149 & 1.91$\times$10$^{11}$ & 2.139 & 3.84$\times$10$^{20}$ & 2.121 & 4.08$\times$10$^{18}$ \\ 
\hline
\textbf{Peakshape} $\delta$ & 0.802 & 1.08$\times$10$^{11}$ & - & 1.202 & 8.88$\times$10$^{14}$ & - & 1.027 & 6.04$\times$10$^{10}$ & 1.915 & 1.29$\times$10$^{19}$ & 1.898 & 2.12$\times$10$^{17}$ \\ 
\hline
\textbf{Peakshape} $\omega$ & 0.791 & 2.60$\times$10$^9$ & - & 1.215 & 4.33$\times$10$^{13}$ & - & 1.011 & 1.42$\times$10$^9$ & 1.954 & 1.19$\times$10$^{18}$ & 1.928 & 1.50$\times$10$^{16}$ \\ 
\hline
\textbf{VHR}       & 0.510 & 4.46$\times$10$^5$ & - & - & - & - & 0.763 & 1.06$\times$10$^7$ & - & - & 1.020 & - \\ 
\hline
\textbf{CGDC}      & \textbf{0.812 (8)} & \textbf{3(1)$\times$10$^{11}$} & \textbf{0.0346 (3)} & \textbf{0.860 (3)} & \textbf{8(2)$\times$10$^{10}$} & \textbf{0.040 (1)} & \textbf{1.120 (3)} & \textbf{3(1)$\times$10$^{11}$} & \textbf{1.068 (3)} & \textbf{1.400 (2)$\times$10$^9$} & - & - \\ 
\hline
\end{tabular}
}
\label{tab:kinetic_parameters}
\end{sidewaystable}

The deconvolution results shown in Figure \ref{Fig5} demonstrate the high potential of numerical fitting methods, which converge towards a solution almost independently of the mathematical model used. Therefore, it is imperative to correlate the mathematical model with other techniques. One of the most clarifying techniques in this context is the $T_{Max}-T_{Stop}$ method.  The $T_{Max}-T_{Stop}$ method (Figure \ref{Fig4}) shows that the mathematical model must be based on at least three components, each with corresponding analytical functions of at least three variables ($E$, $T_m$, $I_m$). If the adjustments to the experimental data are made using linear combinations by adding more expressions, the zones of degeneration increase in size, and the numerical method risks falling into these regions.

Although the $T_{Max}-T_{Stop}$ results show at least three components, the first contribution overlaps. Thus, the mathematical model best linked to the $T_{Max}-T_{Stop}$ results is represented by Equation (\ref{Eq16}), as the contributions related to the low-temperature area show a continuous electron trap structure.

The analysis of the experimental results focuses on the activation energy value $E$ of the prominent peaks, as determined by the computational methods summarized in Table \ref{tab:kinetic_parameters}, followed by the analysis of the frequency factor $s$. Significant disparities have emerged in the $E$ values derived from different methodologies. Acknowledging that the peak shape and VHR methods exhibit exceptionally high sensitivity to overlapping distributions is crucial. Consequently, caution must be exercised when considering the results for peaks 1 and 2 due to this sensitivity. In this case, only a deconvolution study can provide optimal results. However, the $E$ values still show significant differences, indicating that mathematical models are essential for developing an exhaustive deconvolution fitting.

\begin{figure}[h!]
    \centering
    \includegraphics[width=8cm]{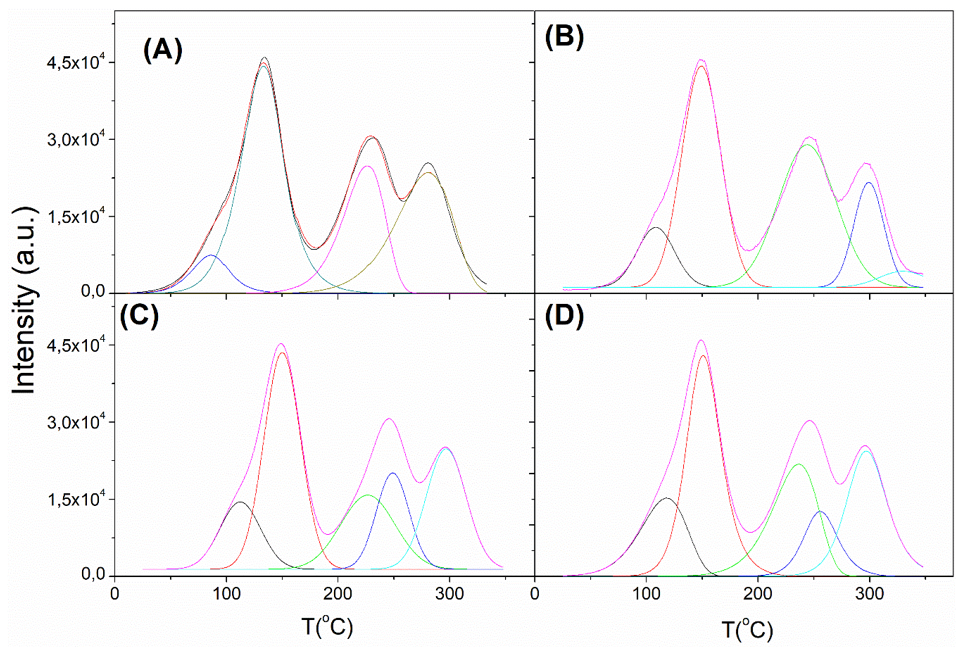} 
    \caption{Fitted experimental TL glow curves of the TL glow peaks of GdAlO$_3$, irradiated with 13.6 Gy of beta radiation, using (A) CGDC, (B) GlowFit, (C) TLAnal, and (D) TLOrigin.}
    \label{Fig5}
\end{figure}

\subsection{Various Heating Rate Method}
Kinetic parameter values obtained using VHR techniques can be strongly altered in TL Glow curves with many overlapping components. This explains the substantial deviation in the activation energy value of peak 3 compared to the average values obtained by other analysis techniques (Table \ref{tab:kinetic_parameters}). Similarly, although less pronounced, it is observed in the values obtained for peak 5 using VHR.

With increasing heating rates, the TL glow curves of undoped GAO phosphor shift to higher temperature values. This shift is illustrated in Figure \ref{Fig1}, which shows the changes in the TL glow curves with varying heating rates. The temperature lag effect, mentioned in numerous studies, explains the discrepancies in the locations of the maximum peak temperatures ($T_M$) \citep{wazir2022computerized, benavente2019thermoluminescence, chen1969glow}. Considering the data, it is essential to account for the temperature lag effect. 

Based on the shape of the TL peak, several expressions have been derived to estimate the activation energy \citep{alajlani2022thermoluminescence, jose2011determination, richhariya2022analysis}. These methods are crucial for obtaining the activation energy $E$ (eV) by considering the position of the glow peak as heating rates increase \citep{depci2021characterization, karmakar2015determination}. These data were used to derive the activation energy and frequency factor values for TL glow signals at both the lowest and highest temperatures.

\section{Discussion}

The TL glow curves of GdAlO$_3$ irradiated with 13.6 Gy of beta radiation have been analyzed using various computational methods to obtain kinetic parameters. The results of these analyses are summarized in Table \ref{tab:kinetic_parameters}.

The deconvolution results reveal that each method offers distinct advantages and reveals different aspects of GAO's TL behavior.

The GlowFit method yields relatively high activation energies ($E$) and frequency factors ($s$) across all peaks. This method is particularly useful for identifying high-energy traps but may overestimate these values due to its sensitivity to overlapping peaks. The high $E$ and $s$ values indicate deep traps with high recombination rates, consistent with the main TL peak observed around 140°C.

The TLAnal method shows significantly lower activation energy for Peak 1 compared to GlowFit, suggesting that it may be better at resolving shallow traps. The TLA method's frequency factors are also quite varied, indicating that this method can more effectively differentiate between traps of different depths. This variability is crucial for understanding GAO's complete trapping and recombination process.

The TLOrigin results are intermediate between GlowFit and TLA, offering a balanced view of shallow and deep traps. This method provides a good compromise, showing moderate activation energies and frequency factors. It demonstrates the importance of using multiple methods to comprehensively understand the TL properties.

The Peak Shape methods ($\tau$, $\delta$, $\omega$) show consistent results for Peak 1, with lower activation energies compared to GlowFit and TLOrigin but higher than TLA. These methods are highly sensitive to the shape of the glow curves and are particularly useful for identifying the symmetry and width of the TL peaks. The consistent results across different peak shape methods indicate the reliability of these techniques in analyzing the symmetry and geometrical properties of the TL peaks.

The VHR method shows the lowest activation energy for Peak 1, indicating its high sensitivity to shallow traps. However, the method's effectiveness diminishes with higher-order peaks due to its reliance on heating rate variations, which can introduce anomalies and errors in the presence of overlapping peaks.

The CGDC method provides a comprehensive analysis by resolving overlapping peaks and distinguishing between continuous and localized trap distributions. The deconvolution results from CGDC indicate multiple contributions to the TL glow curves, with distinct activation energies and frequency factors for each peak. This method's ability to handle complex glow curves makes it particularly suitable for materials like GAO, where multiple trapping sites and recombination processes coexist.

In summary, the results from different methods highlight the complexity of the TL behavior in GAO. The variation in activation energies and frequency factors across methods underscores the need for a multi-faceted approach to TL analysis. The CGDC method stands out due to its robust handling of overlapping peaks and detailed resolution of kinetic parameters. However, each method contributes unique insights, and their combined use provides a more comprehensive understanding of the TL properties of GAO. This comprehensive analysis is crucial for optimizing the material's applications in dosimetry and other fields.

The $T_{Max}-T_{Stop}$ results support these findings by demonstrating at least three distinct contributions, with the first related to a continuous distribution of trapped electrons and the last two corresponding to localized distributions. The sensitivity of peak shape and VHR methods to overlapping distributions necessitates caution when interpreting these results, particularly for peaks 1 and 2. The deconvolution study, particularly through the CGDC method, provides a more accurate and detailed understanding of the activation energies and trapping mechanisms, highlighting the importance of mathematical models in the exhaustive fitting of TL glow curves.

These findings underscore the importance of using a combination of computational methods to fully characterize the TL properties of GAO. By leveraging the strengths of each method, a more accurate and detailed picture of the material's trapping and recombination processes can be achieved, which is essential for its effective application in various scientific fields.

\section{Conclusions}

The study demonstrates that both computational and conventional methods can be effectively applied to analyze individual energy peaks of different kinetic models. By subtracting the residual signal and accurately tracing the initial guess values of kinetic parameters, the fitting of TL glow curves can be optimized for speed and accuracy.

The observed increase in TL intensity with higher heating rates highlights the temperature lag effect. The $T_{Max}-T_{Stop}$ method was employed to understand the behavior of the TL maximum position, showing that its values increased with higher stopping temperatures. This indicates the presence of a quasi-continuous distribution of trapping centers. 

Although a TL Glow curve generated by a continuum of traps can potentially be fitted using a quasi-continuous model, the advantage of using expressions associated with first-order kinetics with continuous trap distributions lies in their comparative simplicity. Quasi-stationary models employ linear combinations of functions with at least three parameters per function, while continuous trap distributions require only a single analytical expression to define their width.

Furthermore, a comprehensive characterization of thermoluminescent materials can be achieved by analyzing their TL glow curves. Estimation of TL kinetic parameters can be performed using various mathematical models, including GlowFit, TLAnal, and the spreadsheet Origin. Among these, the CGCD method emerged as the most suitable technique for the current study. CGCD provided a detailed description of the TL physical process, supported by mathematical models based on a linear combination of functions related to the First Order Kinetic approach.

The study's findings underscore the importance of using a combination of methods to fully characterize the TL properties of GdAlO$_3$ phosphors. The variation in activation energies and frequency factors across different methods emphasizes the need for a multi-faceted approach to TL analysis. The CGCD method, in particular, proved effective in resolving overlapping peaks and providing detailed kinetic parameters, offering significant insights into the trapping and recombination processes within the material.

Overall, this research contributes to a deeper understanding of the TL properties of GdAlO$_3$ phosphors and demonstrates the value of advanced deconvolution techniques in thermoluminescence analysis. Future work should continue to explore and refine these methods, incorporating machine learning algorithms to potentially enhance the accuracy and precision of TL characterization in various scientific and technological fields.

\section*{Acknowledgements}
TRM and RG-S thank IPN Research Project No. 2258. RG-S acknowledges the support provided by SIP20230505-IPN and SIP20240638-IPN, FORDECYT-PRONACES-CONACYT CF-MG-2558591, COFAA-IPN, and EDI-IPN grants.

\bibliography{Teo24Paper}

\bibliographystyle{unsrt}




\end{document}